\documentclass[aps,prb,twocolumn,superscriptaddress,showpacs,amsmath,amssymb,footinbib,longbibliography]{revtex4-2}
\usepackage[english]{babel}
\usepackage{amssymb,amsbsy,amsmath}
\usepackage{graphicx}% Include figure files
\usepackage{dcolumn}% Align table columns on decimal point
\usepackage{bm}% bold math
\usepackage{subfigure}
\usepackage{color}
\usepackage{textcomp}
\usepackage[colorlinks,bookmarks=false,citecolor=blue,linkcolor=red,urlcolor=blue]{hyperref}
\newcommand{\op}[1]{\fontdimen12\textfont3=2pt\fontdimen12\scriptfont3=1.4pt\!\null\mathop{\protect\vphantom{#1}\smash{#1}}\limits_{\sim}\null\!}
\newcommand{\xref}[1]{\protect\ref{#1}}
\newcommand{\figref}[1]{Fig.~\protect\ref{#1}}
\newcommand{\fmref}[1]{(\protect\ref{#1})}
\def\bra#1{\langle \, {#1} \, | \,}
\def\ket#1{\, | \, {#1} \, \rangle}

\renewcommand{\eqref}[1]{Eq.~(\protect\ref{#1})}

\def\erw#1{\langle \, {#1} \, \rangle}

\newcommand{\new}[1]{#1}

\begin{document}
\title{Permanent oscillations and solitary wave behavior in flatband Heisenberg quantum spin systems}

\author{Jannis Eckseler}
\email{jeckseler@physik.uni-bielefeld.de}
\author{J\"urgen Schnack}
\email{jschnack@uni-bielefeld.de}
\affiliation{Fakult\"at f\"ur Physik, Universit\"at Bielefeld, Postfach 100131, D-33501 Bielefeld, Germany}

\date{\today}

\begin{abstract}
\new{Research on the emergence of thermodynamics in closed quantum systems under
unitary time evolution arrived at the consensus that generic systems equilibrate
under rather general assumptions. A new focus of the field is thus on exceptions.} 
Persistent oscillations are \new{one possible} hallmark of non-ergodic time evolution.
While time-crystalline behavior results from, e.g., many-body localization,
here we show that ever-revolving solitary waves emerge in flatband
Heisenberg quantum spin systems. This phenomenon is rather general 
for a variety of frustrated spin systems in one, two, and three dimensions
as well as for Hubbard systems.
\end{abstract}

%\pacs{75.10.Jm,75.50.Xx,75.40.Mg} 

\maketitle

%%%%%%%%%%%%%%%%%%%%%%%%%%%%%%%%%%%%%%%%%%%%%%%%%%%%%%%%%%%%%%%%%%%%%%%%
\section{Introduction}
\label{sec-1}

Ever-lasting oscillations are a fascinating phenomenon 
that got a new twist with the advent 
of time crystals \cite{Wil:PRL12,MBJ:PRB20,HaS:EPL22,RVS:PRR23}, 
where a quantum system exhibits
oscillatory behaviour of some of its observables. 
Although a wider notion of time crystals was introduced in \cite{KMS:A19} 
(compare in particular Fig.~8) the phenomenon has been narrowed
to systems exhibiting many-body localization where a periodic drive 
results in sub-harmonic response.

The example we want to discuss in the following belongs to the class of non-driven 
closed Hamiltonian systems, and thus has got some relationship
with quantum scars \cite{TMA:NP18,HCP:PRL19,MHS:PRB20,KMH:PRB20,PVB:NC21}
as well as Hilbert space fragmentation \cite{KHN:PRB20,Buc:PRL22,MBR:RPP22,WMP:PRL24}.
The appearance of quantum scars, Hilbert space fragmentation, or time crystals
signals non-ergodic/non-thermalizing behavior that contradicts 
our expectation of thermalization that nowadays is the natural
expectation for generic quantum systems even when 
closed, \new{small} and under unitary time evolution
\cite{Deu:PRA91,Sre:PRE94,ScF:NPA96,ScF:PLB97,Tas:PRL98,RDO:N08,Rei:PRL08,PSS:RMP11,ReK:NJP:12,ShF-NJP:12,SKN:PRL14,GoE:RPP16,AKP:AP16,BIS:PR16,STS:QST18,ReG:PA19}.
In order not to interfere with the still evolving notion of time crystals
we prefer to qualify the persistent oscillatory dynamics discussed 
in this article as sufficiently non-trivial (compared to trivial Bloch oscillations 
or single-spin Larmor precessions \cite{Wil:PRL12}). 
Moreover, as we are going to show, it is related to 
the observation of solitary waves and transport in quantum spin systems
\new{and thus has got connections to quantum magnetism \cite{MiS:AP91,ScS:JMMM06}.}

%===================    figure   =================================
\begin{figure}[ht!]
\centering
\includegraphics*[width=0.85\columnwidth]{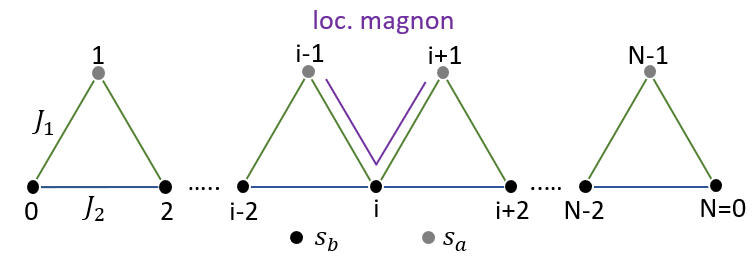}
\includegraphics*[width=0.79\columnwidth]{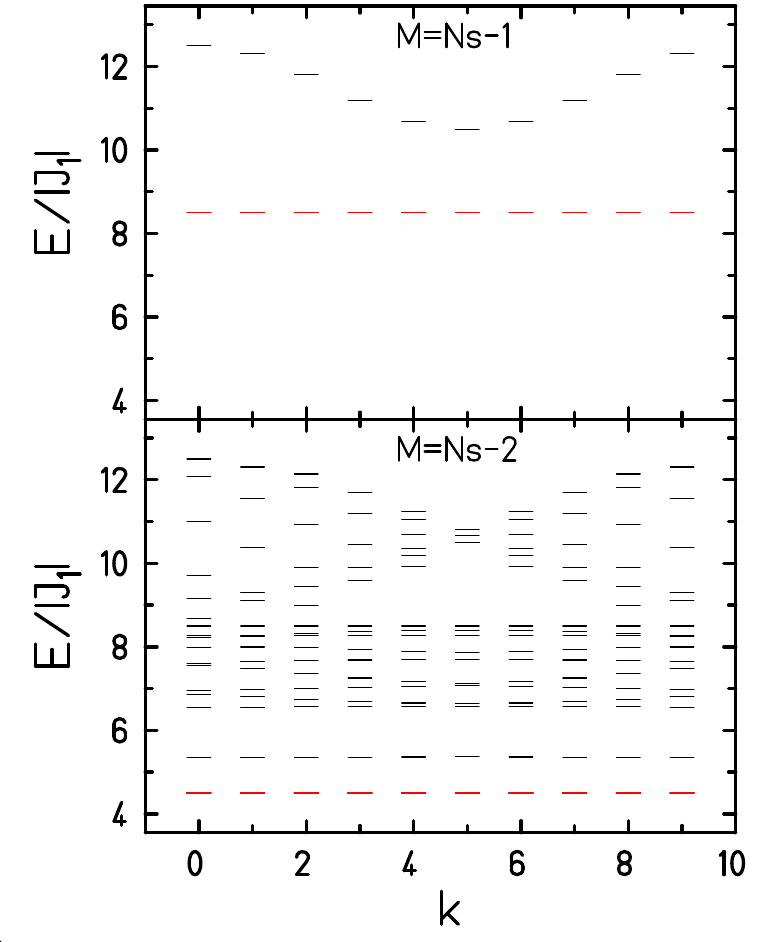}
\caption{Top: Structure of the delta chain with apical spins $s_a$ and basal
spins $s_b$ as well as exchange interactions $J_1$ and $J_2$. 
The spins are numbered $0,1,\dots, N-1$. Periodic boundary conditions
are applied, i.e.\ $N\equiv 0$.
An independent localized one-magnon state is highlighted
that extends over three neighboring sites
as indicated \cite{SSR:EPJB01}.
Bottom: Energy eigenvalues for  
$N = 20$ and $s_a = s_b = \frac{1}{2}$ for one-magnon ($M=N s -1$) and
two-magnon space ($M=N s -2$).
The momentum quantum number $k$ (wave number) runs from $0$ to $N/2-1$,
compare \cite{JES:PRB23}.
}
\label{dispersion}
\end{figure}
%===================    figure   =================================

Magnetic solitons have been detected experimentally in several
magnetic systems \cite{KLH:PRL99,Essler:PRB99,ANI:PRL00,LKG:PB00,NAA:PB01}
for instance as domain-wall or envelope solitons.
From a theoretical point of view magnetic solitons are solutions
of non-linear differential equations as for instance the cubic
Schr\"odinger equation \cite{MiS:AP91,Mik:CSF95}. 
Such non-linear differential equations arise as 
the result of an approximation of the (ordinary) 
time-dependent Schr{\"o}dinger equation, which is linear. For instance, 
a cubic Schr\"odinger equation is obtained when quantum spins are
replaced by a classical spin density \cite{Mat88}.

As discussed already in Ref.~\cite{ScS:JMMM06}, the \emph{linear}
time-dependent Schr\"odinger equation allows for solutions that
move along or across some translationally symmetric 
quantum spin system with frozen shape. 
Such states shall be called solitary waves. More precisely, 
we want to call $\ket{\Psi_s}$ a solitary wave 
if there exists a minimal 
time $\tau>0$ for which the time evolution equals 
(up to a global phase) the shift by one unit cell of the 
spin system \cite{ScS:JMMM06}. A state $\ket{\Psi_s}$ that is a 
superposition of simultaneous eigenstates of energy and momentum 
$\ket{k, E=\gamma k + E_0}, k=0,1,2,\dots$ 
with a linear dispersion relation between the energy eigenvalues $E$ and the momentum 
quantum numbers $k=p/\hbar$ would behave as a solitary wave and move on 
forever on the respective lattice. With periodic boundary conditions as for instance 
naturally given in spin rings this would lead to everlasting revolutions 
around the closed structure.

While in the spectrum of finite-size quantum spin systems 
true linear dispersion relations between more than two energy levels
are typically unlikely, flatband systems give rise to 
\emph{perfectly} 
linear dispersion relations even in non-dense spectra
when combining appropriate multimagnon states 
\cite{MiT:CMP93,SSR:EPJB01,SHS:PRL02,BlN:EPJB03,ZhT:PRB04,DRH:PRB10,DRM:IJMP15,LAF:APX18}.

In order to demonstrate the idea as well as the resulting dynamics 
we choose the one-dimensional delta chain in the Heisenberg model
with spins $s=1/2$ as a model system
which exhibits flat bands in serveral multimagnon subspaces
for a ratio of the two defining exchange interactions 
of $J_2/J_1=1/2$, compare \figref{dispersion}. 
We find it remarkable that flatband spin systems such as the delta chain 
as well as all many other systems such as kagome, square-kagome, pyrochlore etc. 
thus give rise to two rather different phenomena: disorder-free localization
with zero group velocity \cite{MHS:PRB20,JES:PRB23} as well as solitary dynamics
demonstrated in this paper.

The paper is organized as follows. In Section \ref{sec-2} we introduce 
the model, repeat the concepts of independent localized magnons 
\new{as well as} 
flat bands 
and explain how initial states are prepared. Section~\xref{sec-3} demonstrates
numerical examples of the everlasting revolution
\new{and their stability.}
The article closes with a discussion in Section~\ref{sec-4}.

%%%%%%%%%%%%%%%%%%%%%%%%%%%%%%%%%%%%%%%%%%%%%%%%%%%%%%%%%%%%%%%%%%%%%%%%
\section{Essential properties of flatband systems}
\label{sec-2}

The antiferromagnetic delta chain,
also termed sawtooth chain, is shown in \figref{dispersion}~(top).
It is modelled by the Heisenberg model with periodic boundary conditions
%--------------------------------------------------------
\begin{align}
\label{E-2-0}
\op{H} = -2 J_1 \sum_{i=0}^{N-1} \op{\vec{s}}_i \cdot \op{\vec{s}}_{i+1} - 
2 J_2 \sum_{i=0}^{\frac{N}{2}-1} \op{\vec{s}}_{2i} \cdot \op{\vec{s}}_{2i+2} \ .
\end{align}
%--------------------------------------------------------
$\op{\vec{s}}_i$ denotes the spin vector operator at site $i$, 
and $J_1<0$ as well as $J_2<0$
are antiferromagnetic exchange interactions. The unit cell contains two spins 
which gives rise to momentum quantum numbers $k=0, 1, \dots, N/2-1$.
Overall, the eigenstates can be organized according to the present symmetries 
and labeled with total spin $S$, total magnetic quantum number $M$, and 
momentum quantum number (wave number) $k$. The Hilbert space decays into 
mutually orthogonal subspaces according to these quantum numbers. 
In general, the sawtooth chain is not integrable.

In one-magnon space two energy bands appear of which 
one is flat for $J_2/J_1=1/2$, see  \figref{dispersion}~(center). 
This property is equivalent to the existence of 
localized independent one-magnon states (sometimes also termed ``compact localized states" 
\cite{MAP:PRB17,LAF:APX18,CHJ:SB23}) of which one is shown in \figref{dispersion}~(top).
These states can be constructed systematically, compare \cite{SSR:EPJB01,DRM:IJMP15}, as
%--------------------------------------------------------
\begin{eqnarray}
\label{E-2-a}
\ket{loc, i} 
&=& 
\frac{1}{\sqrt{12s_a}}
\left(
\op{s}_{i-1}^{-}
-
2\frac{\sqrt{s_a}}{\sqrt{s_b}}
\op{s}_i^{-}
+
\op{s}_{i+1}^{-}
\right)
\ket{\Omega}
\ ,
\nonumber
\\
\ket{\Omega}
&=&
\ket{m_0=s_b, m_1=s_a, \dots m_{N-1}=s_a}
\ ,
\end{eqnarray}
%---------------------------------------------------------
where $\ket{\Omega}$ is called magnon vacuum. 
The localized independent one-magnon states
are not only eigenstates of the Hamiltonian in one-magnon space, but also (relative)
ground states in this space since they are given by Fourier transforms 
of the respective ground-state flat band in one-magnon space \cite{SSR:EPJB01}.
Out of localized independent one-magnon states one can construct $n$-magnon 
states that are also eigen- and relative groundstates of the Hamiltonian in 
their respective $n$-magnon spaces up to the maximum possible number of
localized independent magnons \cite{SSR:EPJB01,SRM:JPA06}, compare 
$2$-magnon space in \figref{dispersion}~(bottom).
This leads to a strict linear dispersion 
between magnetic quantum number $M$ and ground state energy of 
the ($Ns-M$)-magnon space. 

The desired linear dispersion relation between $E$ and $k$ is then 
obtained by picking appropriate eigenstates $\ket{M, k, \alpha}$
from the respective degenerate ground state manifold. $\alpha$ serves
as a label to enumerate the levels 
\new{within the subspace of degenerate ground states with 
a certain $M$ and $k$.}
To be specific,
%--------------------------------------------------------
\begin{align}
\label{E-2-1}
\ket{\Psi_s} =& c_0 \ket{M=N s, k=0}
\\
+& c_1 \ket{M=N s-1, k=1, \alpha_1}
\nonumber
\\
+& c_2 \ket{M=N s-2, k=2, \alpha_2}
\dots \ ,
\nonumber
\end{align}
%--------------------------------------------------------
where the first state is the magnon vacuum, the second state a
($k=1$)-eigenstate from the flat ground state band in one-magnon space, 
the third a
($k=2$)-eigenstate from the flat ground state band in two-magnon space, and so on.
We consider a superposition of more than two eigenstates as non-trivial
because it is unlikely for generic finite-size 
systems that more than two eigenstates
fulfill a linear dispersion relation \emph{exactly}.

In general, with $\op{U}$ being the time-evolution operator and 
$\op{T}$ the operator that translates (shifts) by one unit cell, 
solitary waves $\ket{\Psi_s}$ fulfill
%--------------------------------------------------------
\begin{align}
\label{E-2-2}
\op{U}(\tau)\ket{\Psi_s}=e^{-i\phi_0}\op{T}^{\pm}\ket{\Psi_s}
\end{align}
%--------------------------------------------------------
for a certain discrete time $\tau$ (up to a global phase).
Inserting decomposition \fmref{E-2-1} yields 
%--------------------------------------------------------
\begin{align}
\label{E-2-3}
E_{\mu}\tau/\hbar = \pm \frac{4 \pi k_{\mu}}{N}
\left(+2\pi m_{\mu} + \phi_0\right) \ , \ m_{\mu} \in \mathbb{Z} 
\end{align}
%--------------------------------------------------------
mentioned above with some arbitrary constants 
in parenthesis due to properties of the complex unit circle. 
This results in a minimal $\tau$ of
%--------------------------------------------------------
\begin{align}
\tau=\frac{\Delta k 4 \pi \hbar}{\Delta E N}
\ ,
\end{align}
%--------------------------------------------------------
were we show $\hbar$ explicitly for convenience \cite{ScS:JMMM06}.
$\Delta k$ and $\Delta E$ are differences according to \fmref{E-2-3}.

To some extent, solitary waves can be shaped depending 
on the number, kind, and amplitude
of its Fourier components, compare \fmref{E-2-1}. 
In general, more Fourier components with a broader range of
momenta yield smaller distributions in space. In addition, 
even for a fixed $k$ several eigenstates can be picked 
since the flat band states are often degenerate. 
However, the construction principle \fmref{E-2-1} acts 
as a limiting condition. In addition, as we are going to see,
the spacial variance of the expectation value of
a local observable depends on the
specific observable and in particular on how many different 
orthogonal subspaces it connects. The operator $\op{s}_{i}^{x}$,
which we are going to use, connects only eigenstates with 
$\Delta M=\pm 1$, which then displays only limited parts of
the solitary wave \fmref{E-2-1}.

In general, one could say that the solitary wave is probed by 
the respective employed operators, and the result shows a stroboscopic 
representation of it. For the same solitary wave such representations 
can be very different. As an example, the operator $\op{s}_{i}^{z}$
would basically show the mean magnetic moment due to symmetry, 
whereas $\op{s}_{i}^{x}$ uncovers the discussed modulation along the 
chain. Higher-order operators would shed light on other details of
the solitary wave.

%%%%%%%%%%%%%%%%%%%%%%%%%%%%%%%%%%%%%%%%%%%%%%%%%%%%%%%%%%%%%%%%%%%%%%%%
\section{Numerical examples}
\label{sec-3}

\subsection{\new{Main result}}
\label{sec-3-1}

\begin{figure}[ht!]
\centering
\includegraphics*[width=0.85\columnwidth]{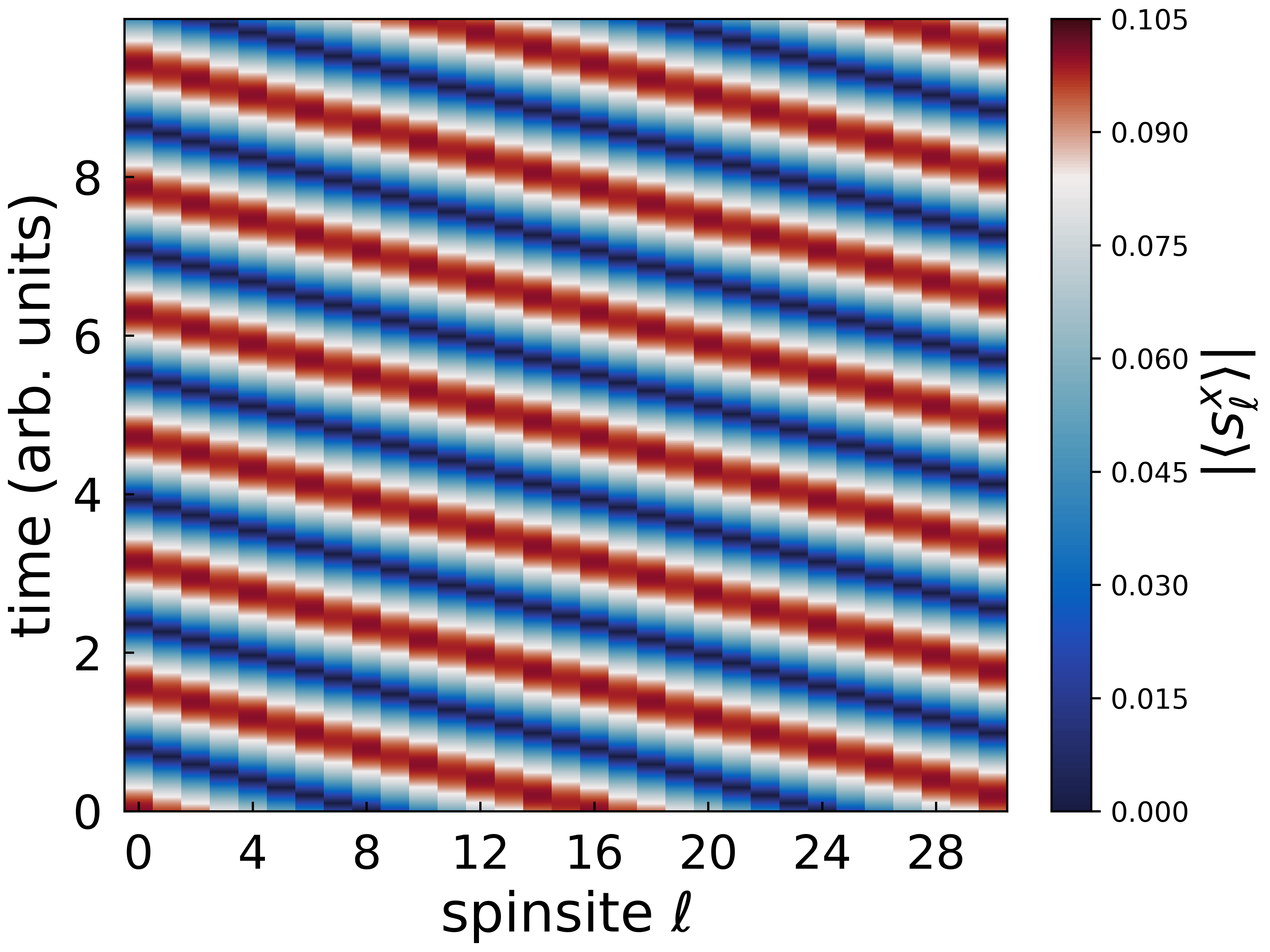}
\caption{Time evolution of the absolute values of expectation values 
$|\erw{\op{s}_{\ell}^x(t)}|$ in a delta chain with N= 32 spins.  
The initial state is a superposition of five eigenstates according to \fmref{E-2-1}.  
The characteristic time $\tau$ can be deduced from the shift by one unit cell 
(of two neighboring spins).
}
\label{flabasolwave-f-2}
\end{figure}

We looked at a delta chain with $N=32$ and $s=1/2$ 
at the flatband point $J_2/J_1=1/2$ where this model is a typical representative 
of strongly frustrated spin systems hosting flat bands. 
Figure \xref{flabasolwave-f-2} shows the time 
evolution of the absolute value of the
expectation value of individual operators $\op{s}_{\ell}^x$
%--------------------------------------------------------
\begin{align}
\label{E-3-1}
\erw{\op{s}_{\ell}^x}
=
\bra{\Psi_s}
\op{s}_{\ell}^x
\ket{\Psi_s}
\end{align}
%--------------------------------------------------------
as a function of time for a superposition of five energy eigenstates
with $k=0,1,2,3,4$. One recognises two features: (1) every individual 
spin expectation value oscillates permanently, and (2) this oscillation 
has got an offset (of size $\tau$) with respect to the neighboring unit cell. 
In total, the picture shows a wave that travels around the delta-chain
with periodic boundary conditions.

%===================    figure   =================================
%===================    figure   =================================

\begin{figure}[ht!]
\centering
\includegraphics*[width=0.85\columnwidth]{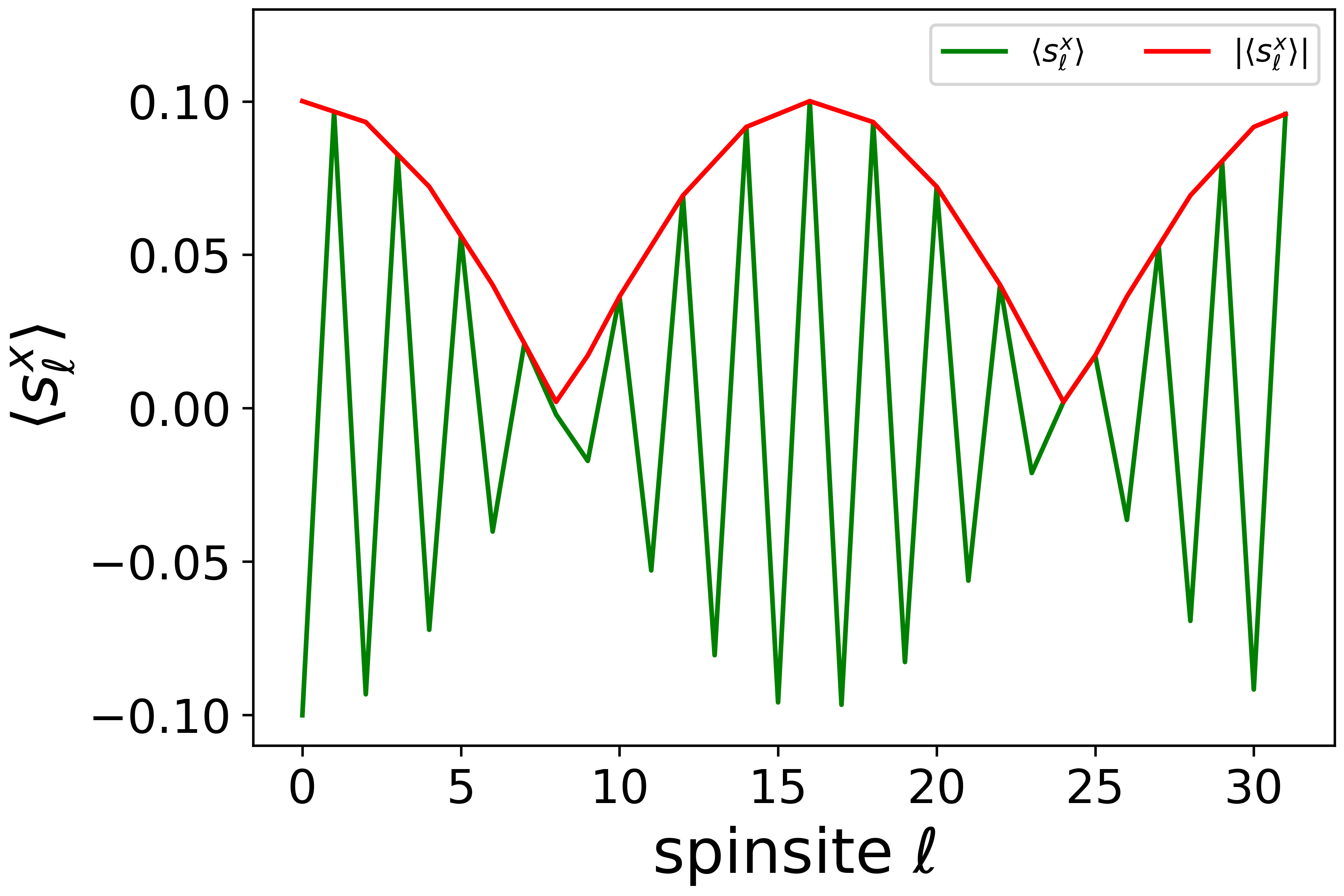}
\caption{$\erw{\op{s}_{\ell}^x}$ and $|\erw{\op{s}_{\ell}^x}|$ 
of an inital solitary wave in a delta chain with $N=32$. 
It can be seen that $|\erw{\op{s}_{\ell}^x}|$ has two peaks over the whole lattice.
}
\label{flabasolwave-f-4-3}
\end{figure}

The pattern seen for a local observable depends on which components this
observable picks out of the superposition. Since the operator $\op{s}_{\ell}^x$
connects only basis states with $\Delta M=\pm 1$ the construction principle  
\fmref{E-2-1} leads to terms in the expectation value that contain only terms
with $\Delta k=\pm 1$ due to the correlation between $M$ and $k$ in \fmref{E-2-1}.
Figure~\xref{flabasolwave-f-4-3} shows the resulting expectation values of 
$\erw{\op{s}_{\ell}^x}$ as well as $|\erw{\op{s}_{\ell}^x}|$ for some moment in time. 
Different operators as well as different construction principles would yield different 
patterns. 

%===================    figure   =================================
%===================    figure   =================================

\new{The perpetual motion visible in \figref{flabasolwave-f-2} can be corroborated 
by inspection of spin currents along the delta chain. We investigated 
%--------------------------------------------------------
\begin{align}
\label{E-3-na}
\op{j}^z
=
-i \sum_{\ell}
\left[\op{H},\op{s}_{\ell}^z\right]
\ ,
\end{align}
%--------------------------------------------------------
and this current is indeed non-zero for the initial states \fmref{E-2-1}
that we investigate. Depending on the $k$-values present in the superposition
\fmref{E-2-1} the current runs clockwise or anticlockwise around the sawtooth
chain with periodic boundary conditions. The overall dynamics can be pictured 
as a screw where the screw moves along the $z$-direction (current) and the
$x$- and $y$-components of the local magnetization rotate according to the
handiness of the screw which can also be inferred from 
\figref{flabasolwave-f-2}. In a broader sense this is related to the observation 
of spin currents in Heisenberg spin systems \cite{ZoP:04,HHB:EPJSP07,BHK:RMP21},
but here it serves to support our claim on the perpetual motion of quantum 
superpositions given by \fmref{E-2-1}.}

\subsection{\new{Stability}}
\label{sec-3-2}

\begin{figure}[ht!]
\centering
\includegraphics*[width=0.85\columnwidth]{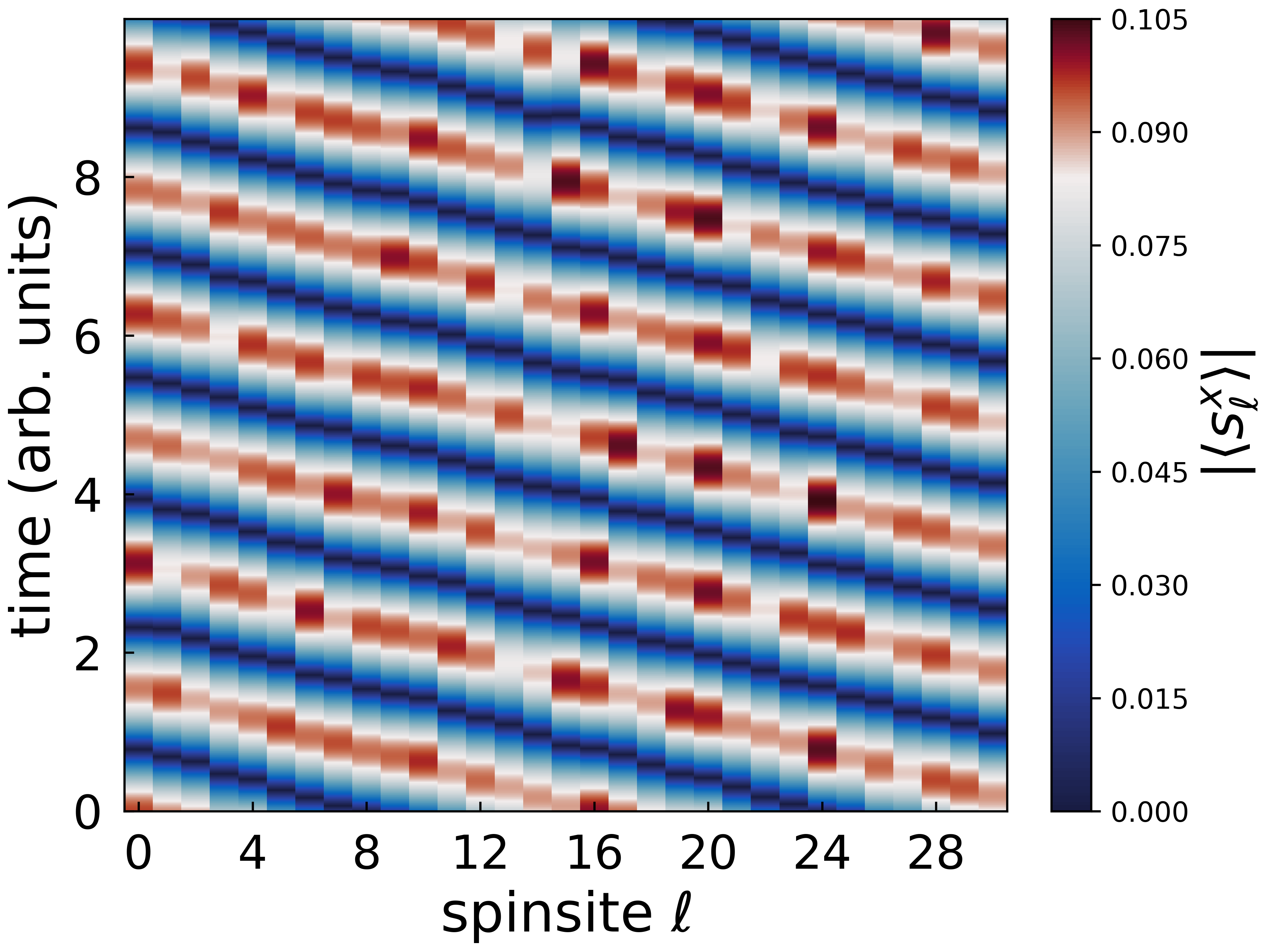}
\caption{Time evolution of the absolute values of expectation values 
$|\erw{\op{s}_{\ell}^x(t)}|$ in a delta chain with N= 32 spins.  
The initial state is a superposition of five eigenstates according 
to \fmref{E-2-1} plus a random Gaussian admixture (each complex entry 
in the representation of a state from the subspace is drawn from a 
Gaussian distribution with mean 0 and standard deviation of 0.45).
}
\label{flabasolwave-f-7-4}
\end{figure}

The flatband point $J_2/J_1=1/2$ defines a fine-tuned Hamiltonian. One could wonder
how robust the dynamics is under disturbances. One possible disturbance concerns the 
initial state \fmref{E-2-1} that might have additional components not compatible with
the construction principle. 
Figure~\xref{flabasolwave-f-7-4} shows the time evolution of an example where 
some random Gaussian contributions have been added to \fmref{E-2-1}. Since the
differential equation is linear these components develop independently of
the perfect state \fmref{E-2-1}. It does appear that the revolving solitary 
contribution is still clearly visible on top of the Gaussian noise.

%===================    figure   =================================
%===================    figure   =================================

\begin{figure}[ht!]
\centering
\includegraphics*[width=0.85\columnwidth]{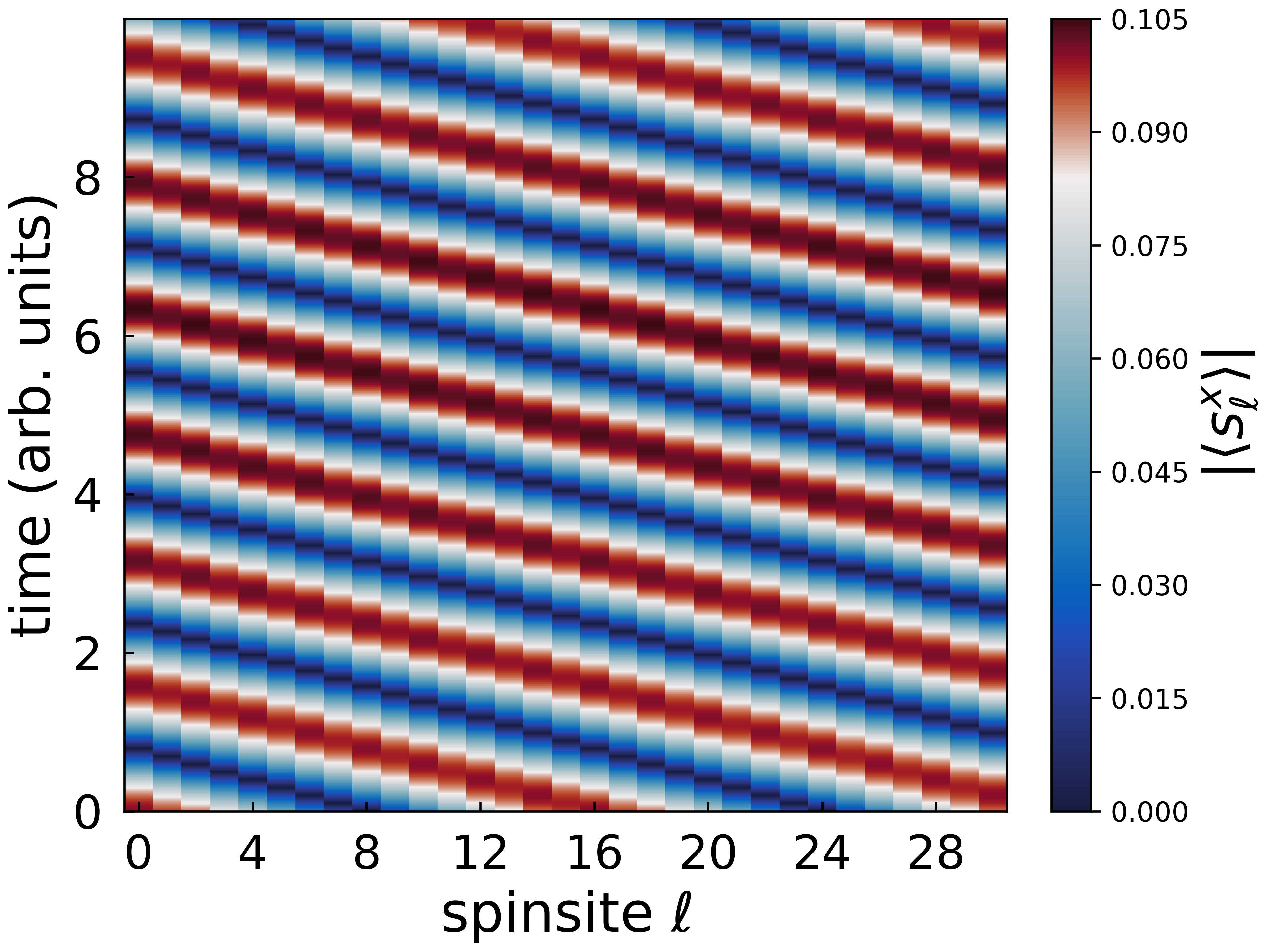}
\caption{Time evolution of the absolute values of expectation values 
$|\erw{\op{s}_{\ell}^x(t)}|$ in a delta chain with N= 32 spins and 
a slightly dispersive band ($J_2/J_1=0.45$).  The initial state is 
a superposition of five eigenstates according to \fmref{E-2-1}.  
}
\label{flabasolwave-f-5}
\end{figure}

Another disturbance consists in moving away from the flat band point.
This situation is covered by \figref{flabasolwave-f-5}, 
where a slightly dispersive 
band was chosen that corresponds to $J_2/J_1=0.45$.
Also here, for a slight disturbance the revolving motion is observable 
for some time. However, one must clearly say that deviations from the 
perfect flat band lead to destruction of periodic motion for longer
times.

%===================    figure   =================================
%===================    figure   =================================

A measure for the stability of the solitary wave is 
the overlap of the shifted state 
with the time evolved one (related to the Jozsa fidelity \cite{Joz:JMO94} 
after one
revolution around the ring)
%--------------------------------------------------------
\begin{align}
\label{E-3-2}
\eta(t)=\bra{\Psi_s} \op{T}^{-1} \op{U}(t) \ket{\Psi_s}
\ ,
\tau/2 \leq t < 3 \tau/2
\ .
\end{align}
%--------------------------------------------------------
We define $\eta(t)$ piece-wise and restart the procedure 
for the next interval accordingly. 
For a perfect solitary wave the absolute value of $\eta(n \tau)$,
$n \in \mathbb{Z}$, is equal to $1$. 

%===================    figure   =================================
\begin{figure}[ht!]
\centering
\includegraphics*[width=0.85\columnwidth]{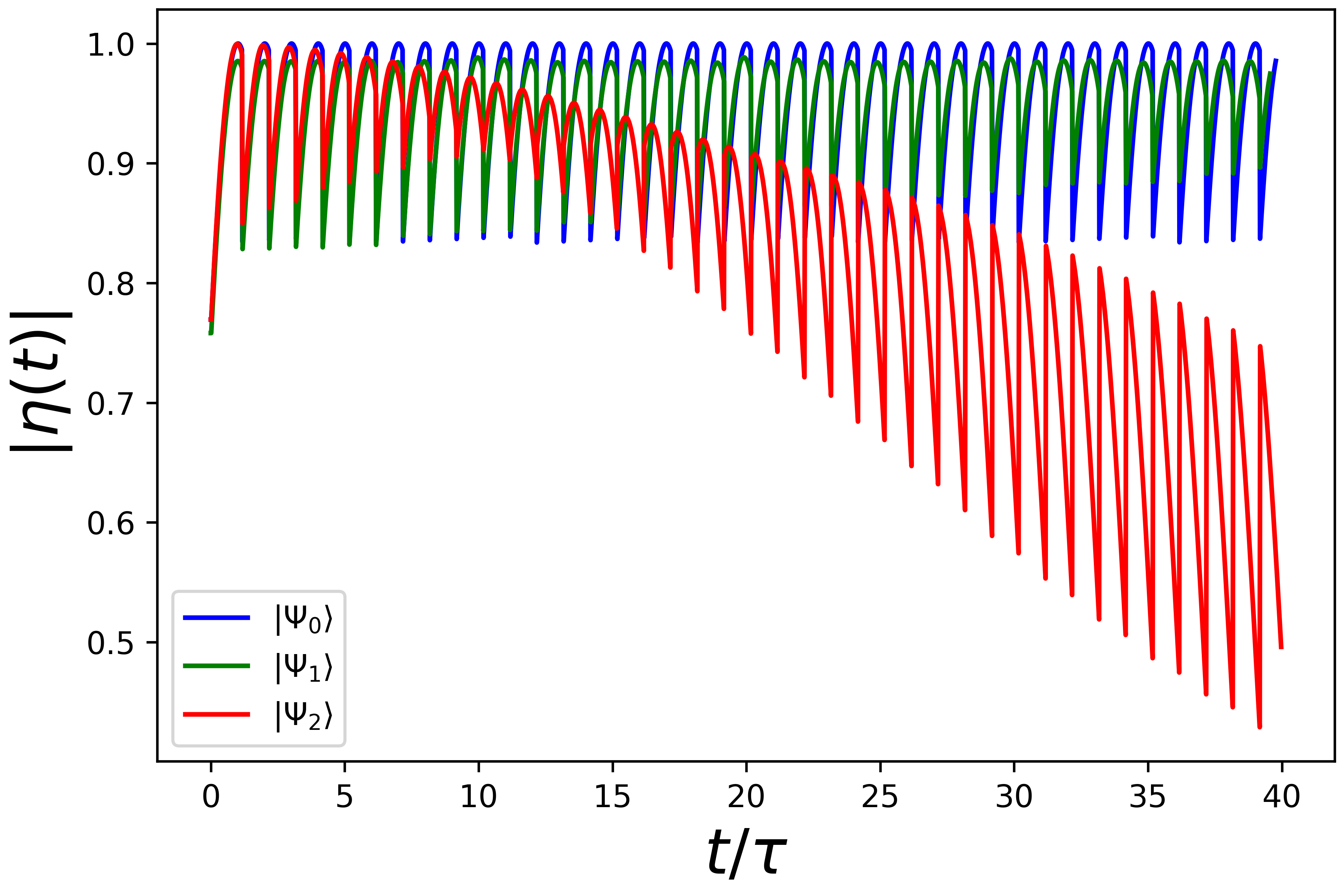}
\caption{Overlap $\eta(t)$ of the shifted state with the time evolved state for 
a perfect solitary wave $\ket{\Psi_0}$ (blue), a solitary wave perturbed with a 
random admixture $\ket{\Psi_1}$ (green) and an approximate solitary 
wave in a system with a slightly dispersive band $\ket{\Psi_2}$ (red).
For the latter case an approximate, effective $\tau$ was used as unit of time.
}
\label{flabasolwave-f-6}
\end{figure}
%===================    figure   =================================

This is seen in \figref{flabasolwave-f-6}
when looking at the blue curve that shows the case of a perfect solitary wave 
$\ket{\Psi_0}=\ket{\Psi_s}$. If some random component is added to a perfect 
solitary wave, resulting in $\ket{\Psi_1}$, 
$|\eta(t)|$ in general will not return to its initial value.
However, $|\eta(t)|$ periodically returns to some smaller value 
since the contribution of the solitary wave develops independently of the 
remainder for the linear Schr{\"o}dinger equation. 
The random component might even equilibrate, i.e., smear out around the ring 
while the solitary-wave contribution still runs unperturbed (green curve),
see discussion in \cite{JES:PRB23}. If the Hamiltonian is slightly off
the flatband scenario, i.e., possesses only dispersive bands, 
an approximate solitary wave $\ket{\Psi_2}$
slowly loses recurrence, and $|\eta(t)|$ decays while still oscillating 
(red curve in \figref{flabasolwave-f-6}).

%===================    figure   =================================
\begin{figure}[ht!]
\centering
\includegraphics*[width=0.85\columnwidth]{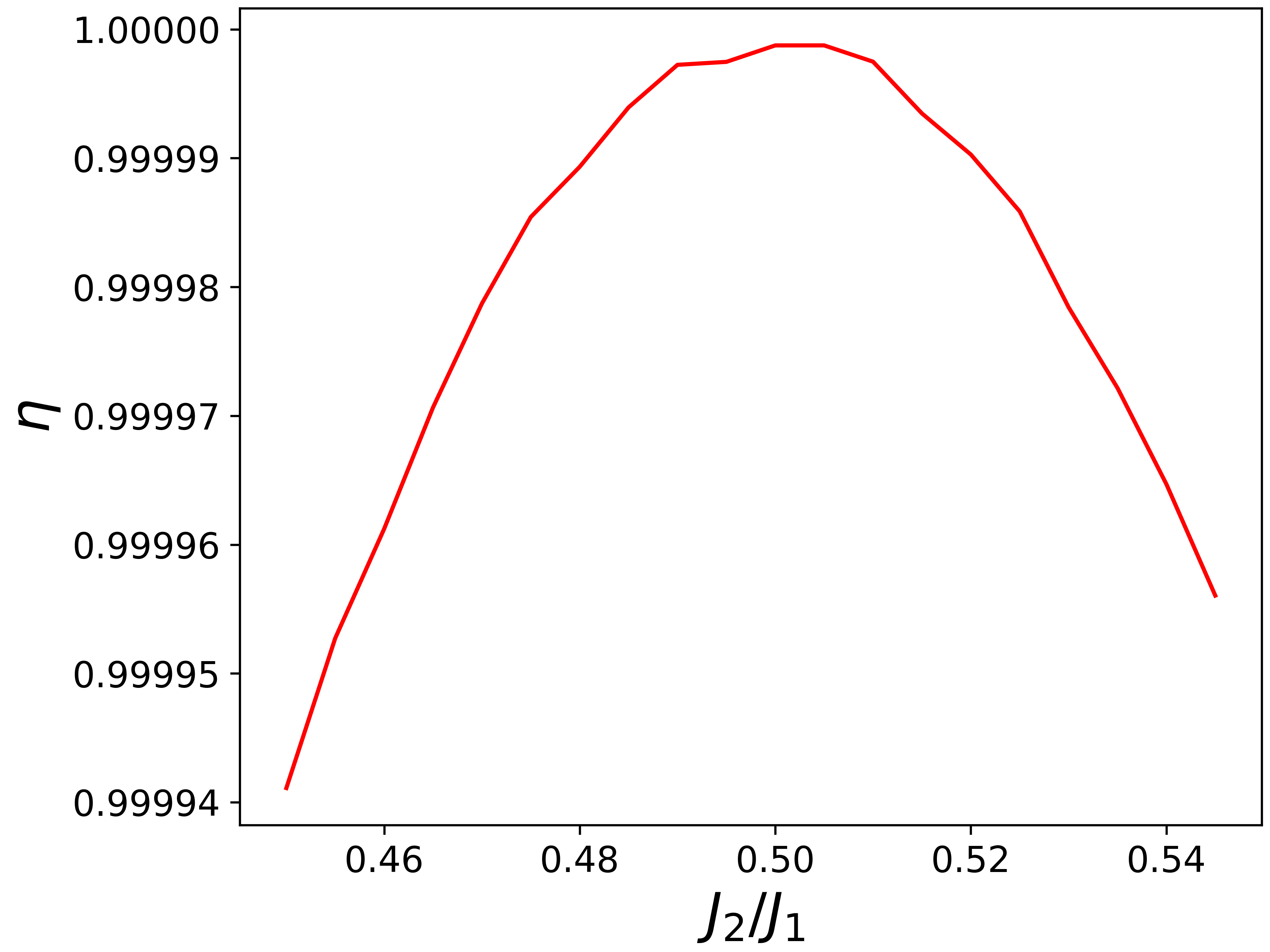}
\caption{\new{Numerical overlap $\eta$ of the shifted state with the time evolved state for 
the next approximate recurrence as function of $J_2/J_1$. The initial state is 
the same as in \figref{flabasolwave-f-2}. For $J_2/J_1=1/2$ the recurrence is 
perfect, the slight deviation from $1.0$ is due to the discretization step of the 
numerical time integration.
For $J_2/J_1\neq 1/2$ $\eta$ stays astonishingly close to one, but of course
drops with detuning away from $J_2/J_1=1/2$.}
}
\label{flabasolwave-f-7}
\end{figure}
%===================    figure   =================================

\new{Figure \xref{flabasolwave-f-7} sketches a stability diagram for the initial 
state used in \figref{flabasolwave-f-2} for various ratios $J_2/J_1$. 
For $J_2/J_1=1/2$ the recurrence is perfect, the slight deviation from $1.0$ 
is due to the discretization step of the numerical time integration that does
not hit the exact recurrence time. 
For $J_2/J_1\neq 1/2$ $\eta$ stays astonishingly close to one, but of course
drops with detuning away from $J_2/J_1=1/2$. This observation means that even 
in cases where the band is not perfectly flat a perpetual motion can be followed
over several cycles.}

%%%%%%%%%%%%%%%%%%%%%%%%%%%%%%%%%%%%%%%%%%%%%%%%%%%%%%%%%%%%%%%%%%%%%%%%
\section{Discussion and conclusions}
\label{sec-4}

In this paper, we demonstrated that certain carefully prepared
initial states of flatband systems give rise to permanent oscillations and solitary wave behavior. We
would like to remind the reader that a magnetic field is not involved. Quantum spin systems with flat 
bands such as the discussed delta chain, the kagome lattice, the square-kagome lattice, 
the pyrochlore lattice
and several other frustrated systems thus do not only show very exciting magnetic properties
like spin-liquid behavior, magnetization plateaus and jumps, they also provide examples 
of non-generic, non-ergodic behavior expressed for instance in disorder-free localization 
with zero group velocity \cite{MHS:PRB20,JES:PRB23} as well as persistent motion of solitary waves.
This very general phenomenon also holds for Hubbard models with flat bands 
\cite{Mie:JPA91A,Mie:JPA91,Mie:JPA92B,Tas:PRL92,MiT:CMP93,Tas:PTP98,BeL:IJMPB13,DRM:IJMPB15,MMS:PRB17,LAF:APX18,TDD:ZN20,DaD:NJP24}.
\new{In a forthcoming publication the phenomenon will be discussed for the technically 
more demanding two-dimensional kagome lattice antiferromagnet \cite{SES:25} based on a
recent thesis \cite{Schlueter:Diss24}.}

Of course, the peculiar dynamics is related to fine-tuned Hamiltonians and sometimes also
fine-tuned initial states. Away from the flatband scenario, strictly permanent oscillations
will not occur, however, depending on the strength of the dispersion 
they might be very long-lived. For further thoughts on the decay of long-lived oscillations 
see \cite{PRA:RSE24}.

Experimentally, the needed initial states could be excited by means of electron paramagnetic
resonance (EPR) at low temperature close to the saturation field. The strong correlation between 
magnetic quantum number and momentum quantum number works in favor of a balanced population
of the needed components.

New developments point at flat bands accessible at small fields. These flat band systems
are characterized by ferromagnetic as well as antiferromagnetic interactions 
\cite{KDN:PRB14,DKR:PRB18,DSD:EPJB20,DmK:JPCM23}
or special XXZ interactions \cite{CPC:PRB19,LMP:PRB20,PSC:PRB21}.

%%%%%%%%%%%%%%%%%%%%%%%%%%%%%%%%%%%%%%%%%%%%%%%%%%%%%%%%%%%%%%%%%%%%%%%%
\section*{Acknowledgment}

This work was supported by the Deutsche Forschungsgemeinschaft DFG
(355031190 (FOR~2692); 397300368 (SCHN~615/25-2)).
We acknowledge support for the publication costs by
the Open Access Publication Fund of Bielefeld University and
the DFG.

%%%%%%%%%%%%%%%%%%%%%%%%%%%%%%%%%%%%%%%%%%%%%%%%%%%%%%%%%%%%%%%%%%%%%%%%
%\bibliographystyle{/home/schnack/tex/sty/revtex4-1/revtex4-1/bibtex/bst/revtex/apsrev4-2}
%\bibliography{js-own.bib,js-other.bib}


\end{document}